\documentclass[prd,aps,showpacs,,nofootinbib,superscriptaddress,preprintnumbers,psf]{revtex4}
\usepackage[english]{babel}
\usepackage{pstricks}
\usepackage{slashed}
\usepackage[dvips]{graphicx}

\newcommand{\gev}{\,{\rm GeV}}

\begin{document}
\date{\today}
 \def\preprint#1{
 }
\preprint{\begin{tabular}{l}
      arXiv \\
      CPHT/S 
    \end{tabular}
 }
%
%
\title{Timelike  Compton scattering with a linearly polarized photon beam.}
\author{ A.T. Goritschnig}
 \affiliation{Institute of Physics - Theory Division, University of Graz, Graz, Austria}
 \affiliation{CPHT, {\'E}cole Polytechnique, CNRS, 91128 Palaiseau, France}
\author{ B. Pire} 
 \affiliation{CPHT, {\'E}cole Polytechnique, CNRS, 91128 Palaiseau, France}
\author{J. Wagner}
 \affiliation{National Centre for Nuclear Research (NCBJ), Warsaw, Poland}

\begin{abstract}
We study the 
timelike  virtual Compton scattering process with a linearly polarized photon beam, in the generalized
Bjorken scaling region and in the medium energy range which will be studied intensely at JLab12 experiments. 
We define new observables and show how they should help to access the polarized quark and gluon generalized parton distributions. 
\end{abstract}
%
\maketitle
\thispagestyle{empty}
\renewcommand{\thesection}{\arabic{section}}

\renewcommand{\thesubsection}{\arabic{subsection}}
\narrowtext
\noindent
\section{Introduction}
The hard exclusive photoproduction of a lepton pair through timelike Compton scattering (TCS)\cite{BDP}
\begin{equation}
\gamma(q_{in}) N(p) \to \gamma^*(q_{out}) N' (p') \to l^-(k)l^+(k')N'(p')\,,
\label{process}
\end{equation}
shares many features with deeply virtual Compton scattering (DVCS)\cite{historyofDVCS}, in the sense that it obeys a factorization property in the generalized Bjorken limit of large $Q^2$ and fixed ratio ${Q^2}/{s}$ with
\begin{equation}
q_{in}^2 =0~~~;~~q_{out}^2 =Q^2 >0 ~~~;~~s=(p+q_{in})^2 \,.
\end{equation} 
In principle this allows one to access the same knowledge of the nucleon structure encoded in universal generalized parton distributions (GPDs)\cite{gpdrev}. The amplitudes of these two reactions are related at Born order by a simple complex conjugation, but they significantly differ at next to leading order (NLO) in the strong coupling constant $\alpha_s$ \cite{NLOTCS,Moutarde:2013qs}. These two processes are the most promising ones to access the three-dimensional partonic picture of the nucleon \cite{Impact} with the help of the extraction of quark and gluon GPDs.
 
The spin structure of the nucleon remains after decades of experimental efforts an open question of hadronic physics; in particular the contribution of gluons stays at a rather unsatisfactory qualitative level, even for integrated parton distributions. Accessing the polarized GPDs  $\tilde H(x, \xi, t)$ and $ \tilde E(x, \xi, t)$ through pseudoscalar meson deep leptoproduction turns out to be more difficult than anticipated. TCS may be a better tool to access them if proper observables are measured.
 
Experimental techniques have recently been developed to yield a linearly polarized, intense photon beam at JLab, using a forward tagger to detect electrons scattered at very small angles at CLAS12 \cite{CLAS12} or using coherent bremsstrahlung technique at the GlueX experiment \cite{Somov:2011zz}. We show here that this will allow new tests of the polarized quark and gluon GPDs.
\section{Kinematics}
We consider the process in which a linearly polarized real photon interacts with an unpolarized proton and produces a lepton pair. To specify the variables we use in the description of this process we  introduce two coordinate systems. The first one, similar to the one introduced in \cite{BDP}, is the $\gamma p$ center­-of-­momentum frame. In the case of a linearly polarized photon we now have a distinguished transverse direction given by the polarization vector, which we choose to point in the $x$-direction:
\begin{equation}
\epsilon(q)^\mu = \delta^{1\mu} \,.
\end{equation}
The particle momenta in that frame are given by:
\begin{eqnarray}
q_{in}^\mu &=& (q^0,0,0,q^0)\,, \nonumber \\
p^\mu &=& (p^0, 0,0,-q^0)\,, \nonumber \\
q_{out}^{ \mu} &=& (q'^{0},\Delta_T \cos\Phi_h,\Delta_T \sin \Phi_h, q'^{3}) \,,\nonumber \\
p'^{ \mu} &=& (p'^{0},-\Delta_T \cos\Phi_h,-\Delta_T \sin \Phi_h, -q'^{3}) \,,
\end{eqnarray}
the scattering angle in that frame reads:
\begin{equation}
\sin \theta_{c.m.} = \frac{|\vec{p}\times \vec{p}~'|}{|\vec{p}||\vec{p}~'|} = \frac{\Delta_T}{|\vec{p}~'|} \sim \mathcal{O}\left(\sqrt{\frac{-t}{s}}\right) \,,
\end{equation}
where the squared momentum transfer $t$ is defined as $t \equiv \Delta^2 = (p'-p)^2$, and $\Phi_h$ is the angle between the polarization vector and the hadronic plane:
\begin{equation}
\sin \Phi_h = \vec{\epsilon}(q_{in})\cdot \vec{n} =  \vec{\epsilon}(q_{in})\cdot \frac{\vec{p}~'\times \vec{p}}{|\vec{p}~'\times \vec{p}|}\,,
\end{equation}
where $\vec{n}$ is the vector normal to the hadronic plane.

Two additional variables $\theta$ and $\phi$ are defined, as in \cite{BDP}, as the polar and azimuthal angles of $\vec{k}$ in the $l^+l^-$ center-of-momentum frame with the $z$-axis along $-\vec{p}~'$, and the other axes such that $\vec{p}$ lies in the $x$-$z$ plane and has positive $x$ component. 

The  full Compton scattering amplitude in its factorized form is conveniently expressed through Compton Form Factors $\mathcal{H}, \mathcal{E},\tilde{\mathcal{H}}, \tilde{\mathcal{E}}$:
\begin{eqnarray}
\mathcal{A}^{\mu\nu}(\eta,t) &=& - e^2 \frac{1}{(p+p')\cdot n_-}\, \bar{u}(p^{\prime}) 
\bigg[\,
g_T^{\mu\nu} \, \Big(
      {\mathcal{H}(\eta,t)} \, \slashed n_- +
      {\mathcal{E}(\eta,t)} \, \frac{i \sigma^{\alpha\rho}n_{-\alpha}\Delta_{\rho}}{2 M}
   \Big) \nonumber \\
& & \phantom{AAAAAAAAA}
   +i\epsilon_T^{\mu\nu}\, \Big(
    {\tilde{\mathcal{H}}(\eta,t)} \, \slashed n_-\gamma_5 +
      {\tilde{\mathcal{E}}(\eta,t)} \, \frac{\Delta^{+\alpha}n_{-\alpha}\gamma_5}{2 M}
    \Big)
\,\bigg] u(p) \,,
\label{eq:amplCFF}
\end{eqnarray}
where :
\begin{eqnarray}
\left\{\mathcal{H}, \mathcal{E}\right\}(\eta,t) &=& + \int_{-1}^1 dx \,
\left(\sum_q T^q(x,\eta)\left\{H^q,E^q\right\}(x,\eta,t)
 + T^g(x,\eta)\left\{H^g,E^g\right\}(x,\eta,t)\right) \nonumber \\
\left\{\tilde \mathcal{H}, \tilde \mathcal{E}\right\}(\eta,t) &=& - \int_{-1}^1 dx \,
\left(\sum_q \tilde T^q(x,\eta)\left\{\tilde H^q,\tilde E^q\right\}(x,\eta,t) 
+\tilde T^g(x,\eta)\left\{\tilde H^g,\tilde E^g\right\}(x,\eta,t)\right).
\label{eq:CFF}
\end{eqnarray}
The renormalized coefficient functions $T^{q,g}, \tilde{T}^{q,g}$ calculated at LO and NLO in $\alpha_S$ can be found in \cite{Moutarde:2013qs}; we use standard definitions of GPDs $H^{q,g}, \tilde{H}^{q,g},E^{q,g}, \tilde{E}^{q,g} $ as given in Diehl's review \cite{gpdrev}. One has to remember that in the case of the $\gamma p$ center-of-momentum frame defined above, $\vec{p}$ points in the negative $z$-direction, so $n^{+\mu} = \frac{1}{\sqrt{2}}(1,0,0,-1)$, $n^{-\mu} = \frac{1}{\sqrt{2}}(1,0,0,1)$. The scaling variable $\eta$ (so-called skewness variable) is defined as:
\begin{eqnarray}
\eta &\equiv& -\frac{(q_{in}-q_{out})\cdot(q_{in}+q_{out})}{(p+p')\cdot(q_{in}+q_{out})} \approx \frac{Q^2}{2s-Q^2} \,.
\end{eqnarray}
In equations (\ref{eq:amplCFF}, \ref{eq:CFF}) we have suppressed dependencies on $Q^2$, the factorization scale $\mu_F$, and the renormalization scale $\mu_R$ in the notation. Throughout the whole paper we use $Q=\mu_F = \mu_R$.
\section{Cross sections and observables}

\begin{figure}
\includegraphics[keepaspectratio,width=0.45\textwidth,angle=0]{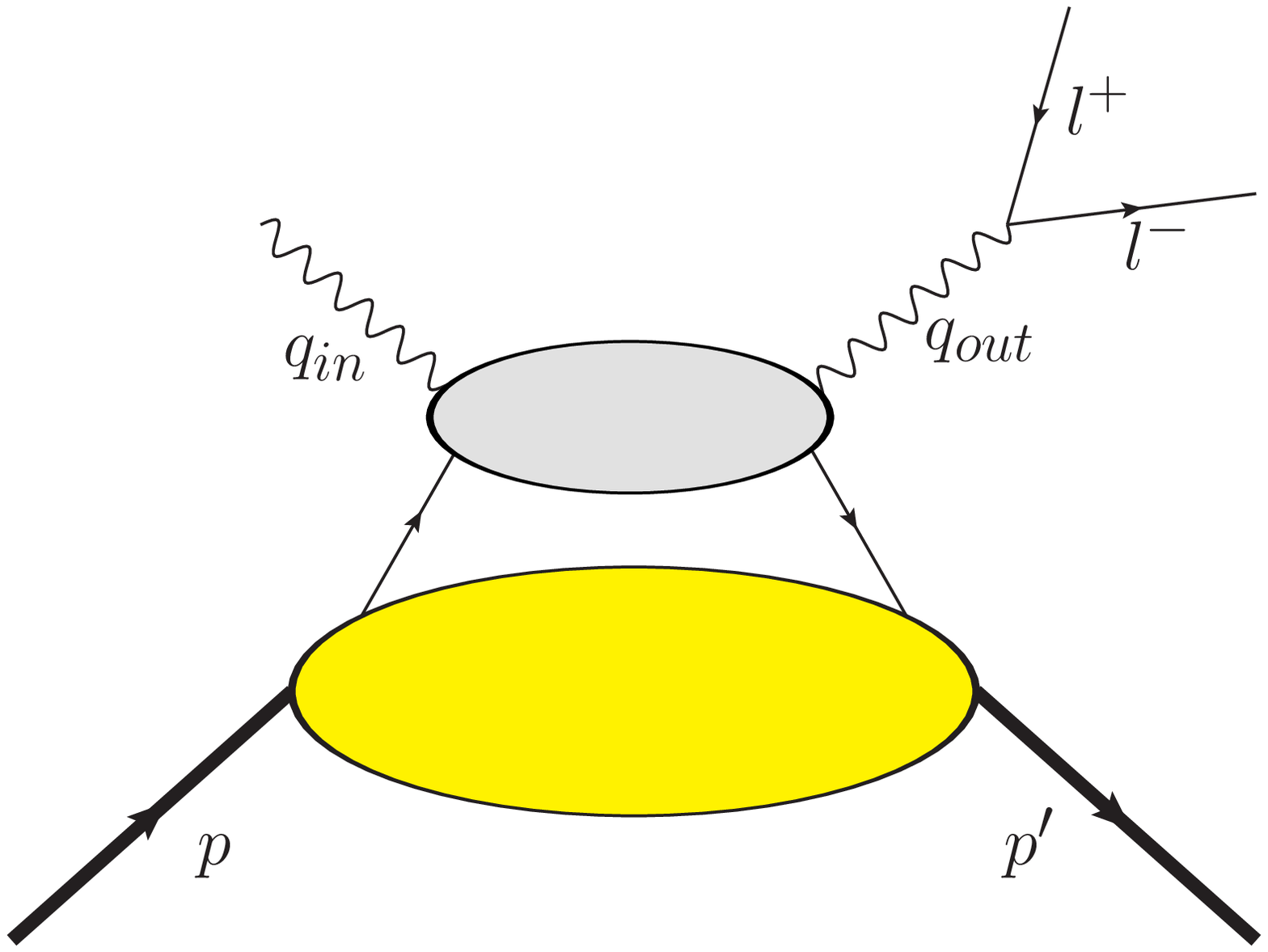} 
\hspace{0.05\textwidth}
\includegraphics[keepaspectratio,width=0.35\textwidth,angle=0]{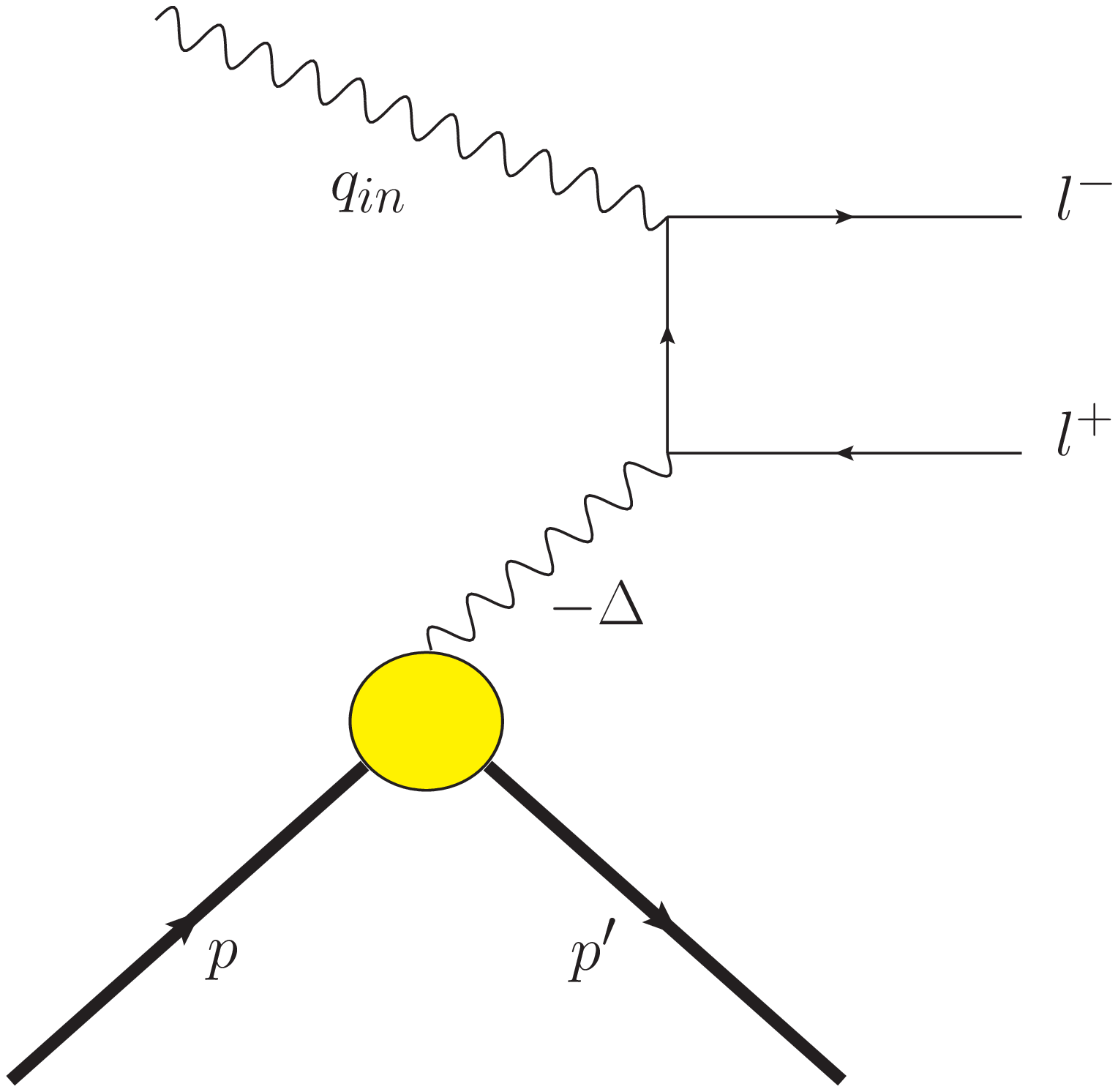} 
\caption{Two mechanisms contributing to the process of photoproduction of lepton pairs: timelike Compton Scattering (Left) and Bethe-Heitler (Right).}
\label{fig:BH_TCS}
\end{figure}

The physical process in which we would like to study TCS is the exclusive photoproduction of a lepton pair. 
But to this process it is not only TCS which contributes, there is also a Bethe-Heitler (BH) mechanism. 
There the incoming photon fluctuates into the lepton pair, which then goes on-shell 
via the exchange of a virtual photon with the proton, cf. figure \ref{fig:BH_TCS}. 
As opposed to DVCS the BH contribution even dominates over the TCS one in the whole kinematical range under consideration. 
The simultaneous presence of the TCS and the BH mechanism at the amplitude level 
leads to the pleasant fact of quantum mechanical interference when considering cross sections. 
Namely, it is this interference contribution which will allow us to extract information on the Compton process itself. 
The interference part can be accessed through the study of the angular distribution of the produced lepton pair. 
In the following we will present our predictions for differential $\gamma p \to l^+ l^- p$ cross sections with polarized incoming photons. 
When doing that we average over the initial proton and sum over all final particle polarizations.  \\ 

Up to the leading order accuracy the contribution of the TCS mechanism to the differential $\gamma p \to l^+ l^- p$ photoproduction cross section for fixed polarization of the incoming photon reads:
\begin{eqnarray}
\frac{d\sigma^{(TCS)}}{dQ^2 dt d\Omega_{l^+l^-} d \Phi_h} 
& \,=\,  & \frac{1}{2^{11}\pi^5}\frac{1}{s^2}\cdot \frac12 \,\sum\, \mid M_{TCS} (\epsilon) \mid^2 \nonumber\\
& \, =\,  & 
\frac{\alpha^3}{16\pi^2s^2}\frac{1}{Q^2}\left( 1 - \sin^2\theta \cos^2(\Phi_h + \phi ) \right) 
\left[ (1-\eta^2) \mid\mathcal{H}\mid^2 - (\eta^2+\frac{t}{4M^2}) \mid\mathcal{E}\mid^2 - 2\eta^2\text{Re}(\mathcal{H^*}\mathcal{E}) \right] \nonumber\\
&     +    &  
\frac{\alpha^3}{16\pi^2s^2}\frac{1}{Q^2}\left( 1 - \sin^2\theta \sin^2(\Phi_h + \phi ) \right) 
\left[ (1-\eta^2) \mid\tilde{\mathcal{H}}\mid^2 - \eta^2\frac{t}{4M^2} \mid\tilde{\mathcal{E}}\mid^2 - 2\eta^2\text{Re}(\tilde{\mathcal{H}}^*\tilde{\mathcal{E}}) \right]\,,  
\label{eq:sigmaX-TCS}
\end{eqnarray}
where $d\Omega_{l^+l^-} = d(\cos\theta) ~d\phi$.

We shall not write down explicitely here the BH contribution, it should only be noted that this contribution is parametrically enhanced by a factor of $Q^{2}/(-t)$ as compared to the TCS part of the differential cross section. 
Let us now turn to the contribution of the interference between the TCS and BH mechanisms 
to the differential $\gamma p \to l^+ l^- p$ cross section. 
For incoming photons with polarization vector as in Eq.~(3) it, up to our accuracy, reads:
\begin{eqnarray}
\frac{d\sigma^{(INT)}}{dQ^2 dt d\Omega_{l^+l^-} d \Phi_h} 
& \,=\,  &  \frac{1}{2^{11}\pi^5}\frac{1}{s^2}\cdot\frac12 \,\sum\, \bigg( M_{TCS} (\epsilon)M_{BH}^* (\epsilon) + c.c.  \bigg) \nonumber\\
& \, \equiv\,  & 
\frac{d\sigma^{(INT)}_{unpol}}{dQ^2 dt d\Omega_{l^+l^-} d \Phi_h} +
\frac{d\sigma^{(INT)}_{linpol}}{dQ^2 dt d\Omega_{l^+l^-} d \Phi_h} \,,
\end{eqnarray}
where:
\begin{eqnarray}
\frac{d\sigma^{(INT)}_{unpol}}{dQ^2 dt d\Omega_{l^+l^-} d \Phi_h}
& \, =\,  & \phantom{-}
\frac{\alpha^3}{16\pi^2s^2}\frac{1}{Q^2} \left(\frac{4s\mid{\Delta}_\perp\mid}{Qt}\right)\,
\bigg( \frac{1 + \cos^2\theta}{\sin\theta}\cos\phi \bigg) 
 \text{Re}\left[\mathcal{H}F_1 - \frac{t}{4M^2} \mathcal{E}F_2 - \eta\tilde{\mathcal{H}}(F_1 + F_2)) \right]  \,,\nonumber\\
\frac{d\sigma^{(INT)}_{linpol}}{dQ^2 dt d\Omega_{l^+l^-} d \Phi_h} 
& \, =\,  & -  
\frac{\alpha^3}{16\pi^2s^2}\frac{1}{Q^2} \left(\frac{4s\mid{\Delta}_\perp\mid}{Qt}\right) \,  
           \bigg(\sin\theta\cos(2\Phi_h + 3\phi) \bigg) 
\text{Re}\left[\mathcal{H}F_1 - \frac{t}{4M^2} \mathcal{E}F_2 + \eta\tilde{\mathcal{H}}(F_1 + F_2)) \right] \,.
\label{eq:sigmaX-INT}
\end{eqnarray}
%
%
The extraction of information from the interference part through the study of 
the angular distribution of the lepton pair relies on the following observation: if one reverses the lepton charge, which is equivalent to an exchange $l \leftrightarrow l^\prime$, the TCS and the BH amplitude transform with opposite signs. 
Thus, the cross sections associated with the TCS and the BH mechanism transform even under this exchange, but the interference cross section is odd. 
If one now considers observables which also change sign when exchanging $l$ and $l^\prime$, they give direct access to the interference term. 

We define two observables sensitive to the unpolarized and linearly polarized part of interference cross section. First one is similar (up to the terms formally of the order $t/Q^2$ or $M^2/Q^2$) to the R ratio defined in \cite{BDP} in the case of an unpolarized photon beam :
\begin{equation}
\tilde{R} = 
\frac{\int_0^{2\pi} d\Phi_h 2\int_0^{2\pi} d\phi\cos(\phi) \int_{\pi/4}^{3\pi/4}\sin \theta d \theta\frac{d \sigma}{dt dQ^2 d\Omega d\Phi_h}}
{\int_0^{2\pi} d\Phi_h \int_0^{2\pi} d\phi \int_{\pi/4}^{3\pi/4}\sin \theta d \theta\frac{d \sigma}{dt dQ^2 d\Omega d\Phi_h}}\,.
\end{equation}
The second observable projects out the $d\sigma_{linpol}^{(INT)}$ part of the interference cross section:
\begin{equation}
\tilde{R}_3 = 
\frac{2\int_0^{2\pi} d\Phi_h \cos(2\Phi_h) 2\int_0^{2\pi} d\phi\cos(3\phi) \int_{\pi/4}^{3\pi/4}\sin \theta d \theta\frac{d \sigma}{dt dQ^2 d\Omega d\Phi_h}}
{\int_0^{2\pi} d\Phi_h \int_0^{2\pi} d\phi \int_{\pi/4}^{3\pi/4}\sin \theta d \theta\frac{d \sigma}{dt dQ^2 d\Omega d\Phi_h}}\,.
\end{equation}
Making use of $\tilde{R}$ and $\tilde{R}_3$ we can define the following observable which is sensitive only to the interference term and which provides us with information about $\tilde{\mathcal{H}}$:
\begin{equation}
C =  \frac{\tilde{R}}{\tilde{R}_3} =
\frac{2-3\pi}{2+\pi}
\frac
{
	 \text{Re}\left[\mathcal{H}F_1 - \frac{t}{4M^2} \mathcal{E}F_2 - 	\eta\tilde{\mathcal{H}}(F_1 + F_2)\right]
}
{
	 \text{Re}\left[\mathcal{H}F_1 - \frac{t}{4M^2} \mathcal{E}F_2 + 	\eta\tilde{\mathcal{H}}(F_1 + F_2)\right]
}\,.
\end{equation}

\begin{figure}
\includegraphics[keepaspectratio,width=0.45\textwidth,angle=0]{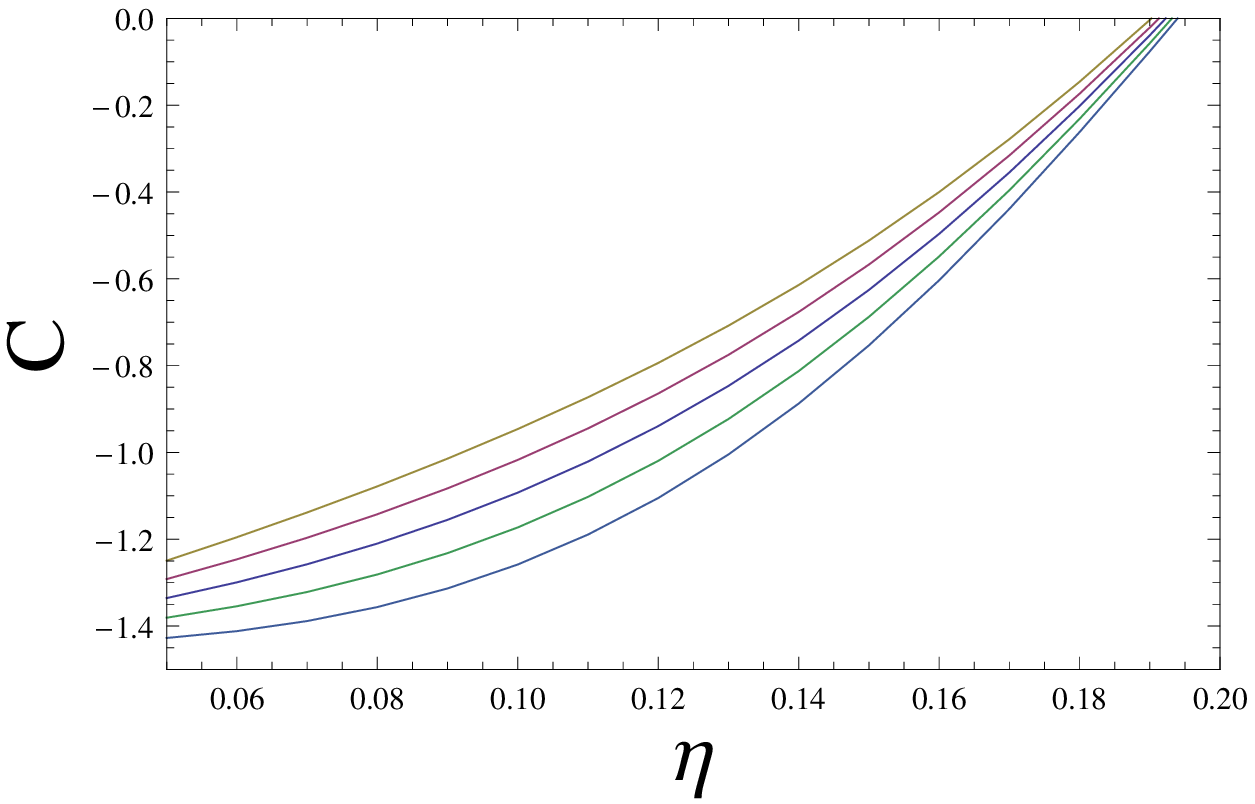} 
\hspace{0.05\textwidth}
\includegraphics[keepaspectratio,width=0.45\textwidth,angle=0]{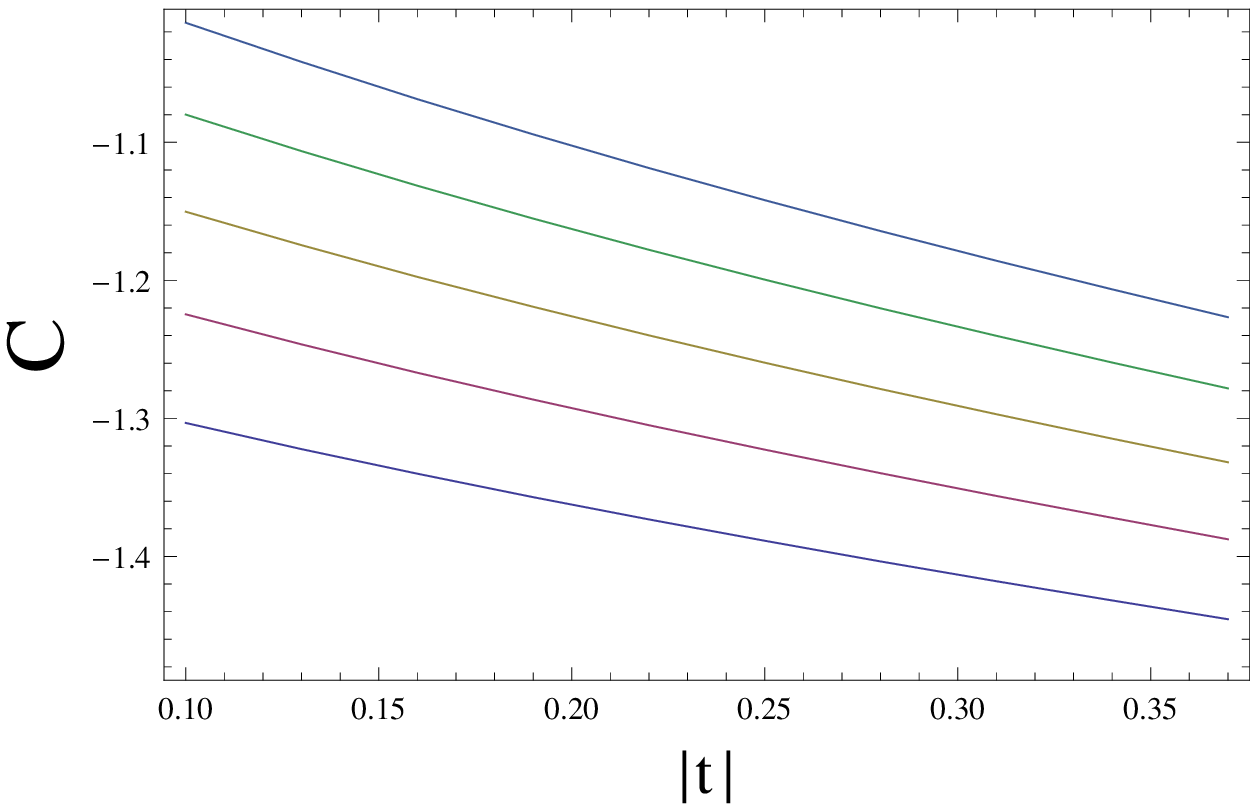} 
\caption{(Left) $C$ as a function of $\eta$ calculated for $Q^2=4 \gev^2$ and $t= t_0(\eta)$. (Right) $C$ as a function of $t$ for  $Q^2=4 \gev^2$ and $\eta= 0.1$. Different curves correspond to different polarized GPDs $\tilde{H}_g = \{-1,0,1,2,3\} \cdot \tilde{H}_g^{GK}$. Calculations where perfomed with NLO accuracy and $\alpha_S=0.3$.}
\label{fig:tildeC}
\end{figure}

In order to illustrate the dependence of the proposed observable on $\tilde{H}$ we make use of the Goloskokov-Kroll GPDs model based on fits to Deeply Virtual Meson Production \cite{GK}. A detailed phenomenological analysis based on that model may be found for example in \cite{KMS}. We present our numerical estimates with the GPD $E$ set to zero, as it is largely unknown and also suppressed by the small kinematic factor $\frac{t}{4M^2}$. 
On the left-hand side of figure \ref{fig:tildeC} we present the ratio $C$ as a function of $\eta$, calculated for $Q^2 = 4 \gev^2$ and a minimal value of $t= t_0(\eta) = -4\eta^2 M^2/(1-\eta^2)$. To study the sensitivity of $\eta$ on the GPD $\tilde{H}$ we have varied the gluonic contribution, taking it as $\tilde{H}_g = \{-1,0,1,2,3\}\cdot \tilde{H_g}^{GK}$, where $\tilde{H_g}^{GK}$ is given by the Goloskokov model. On the right-hand side of figure \ref{fig:tildeC} we show $C$ as a function of $t$ for $Q^2 = 4 \gev^2$ and $\eta =0.1$. In both cases significant effects of the order of $30\%$ are visible proving that $C$ is a good observable sensitive to $\tilde{H}$.  
\section{Conclusions}
We have investigated the photoproduction of lepton pairs off a proton in the generalized Bjorken limit where we have considered the incoming photons to be linearly polarized. To this lepton-pair production a timelike Compton-scattering and a Bethe-Heitler process contribute. When studying their interference one gets direct access to the amplitude of the Compton process which is our main interest. Through it one gets information on the nucleon structure in terms of generalized parton distributions. To this aim we have calculated at leading twist and next to leading $\alpha_S$ order the linearly polarized photoproduction cross sections for the timelike Compton scattering, the Bethe-Heitler process as well as for their interference contribution (where we have neglected  QED radiative corrections \cite{Akushevich:2014zha}). After introducing new observables which are sensitive to different angular dependencies of the outgoing leptons we have been able to project out cross-section contributions associated with particular generalized parton distributions. These observables, which open new and interesting possibilities to study generalized parton distributions, should also be measurable in line with the future experimental program at JLab.

\section*{Acknowledgements} 
We are grateful to Markus Diehl, Charles Hyde, Herv\'e Moutarde, Pawel Nadel-Turonski, Franck Sabati\'e, Stepan Stepanyan, Lech Szymanowski for useful discussions and correspondance. 

This work is partly supported by the Polish NCN Grant No. DEC-2011/01/D/ST2/02069 and the Austrian FWF Grant No. J 3163-N16.


\end{document}